\documentclass[journal]{IEEEtran}
\usepackage{stfloats}
\usepackage[dvips]{graphicx}
\graphicspath{{../eps/}{../pdf/}{../jpeg/}}
\DeclareGraphicsExtensions{.eps,.pdf,.jpeg,.png}
\usepackage{cite}
\usepackage[cmex10]{amsmath}
\usepackage{amsthm, amsfonts, mdwmath, mdwtab, eqparbox, psfrag, setspace}
\usepackage[tight,footnotesize,sf,SF]{subfigure} 
\usepackage{multicol, blindtext}
\usepackage{overpic}
\usepackage{url} 
\usepackage{color}
\usepackage{tikz}
\usepackage{lipsum}
\definecolor{dg}{rgb}{0.1,0.55,0.15}

\usepackage{algorithm}
\usepackage{algpseudocode}
\newtheorem{proposition}{Proposition}

\usepackage{fancyhdr}
\usepackage{acronym}
\usepackage{flushend}
\usepackage{cancel}
\usepackage[T1]{fontenc}

\usepackage[normalem]{ulem}
\newcommand\bout{\bgroup\markoverwith
{\textcolor{blue}{\rule[.5ex]{2pt}{1.0pt}}}\ULon}
\usepackage{xcolor}

\begin{document}

\setlength{\abovedisplayskip}{4pt}
\setlength{\belowdisplayskip}{4pt}
\setlength{\parskip}{0pt}

\acrodef{BS}[BS]{base station}
\acrodef{NE}[NE]{Nash equilibrium}
\acrodef{PPP}[PPP]{Poisson point process}
\acrodef{MNO}[MNO]{mobile network operator}
\acrodef{SDC}[SDC]{strict diagonal concave}
\acrodef{MPP}[MPP]{Mat\'ern-hardcore point process}

\title{Spectrum Allocation for Multi-Operator Device-to-Device Communication} 
\author{
\IEEEauthorblockN{Byungjin Cho\IEEEauthorrefmark{1}, Konstantinos Koufos\IEEEauthorrefmark{1}, Riku J\"antti\IEEEauthorrefmark{1}, Zexian Li\IEEEauthorrefmark{2}, and Mikko A. Uusitalo\IEEEauthorrefmark{2}} \\
\IEEEauthorblockA{\IEEEauthorrefmark{1}Communications and Networking Department, Aalto University, Finland
\\Email:\{Byungjin.Cho, Konstantinos.Koufos, Riku.Jantti\}@aalto.fi} \\
\IEEEauthorblockA{\IEEEauthorrefmark{2}Nokia Technologies, Espoo, Finland \\
Email:\{Zexian.Li, Mikko.A.Uusitalo\}@nokia.com}\vspace*{-.55cm}}

\maketitle

\begin{abstract}
In order to harvest the business potential of device-to-device (D2D) communication, direct communication between devices subscribed to different mobile operators should be supported. This would also support meeting requirements resulting from D2D relevant scenarios, like vehicle-to-vehicle communication. In this paper, we propose to allocate the multi-operator D2D communication over dedicated cellular spectral resources contributed from both operators. Ideally, the operators should negotiate about the amount of spectrum to contribute, without revealing proprietary information to each other and/or to other parties. One possible way to do that is to use the sequence of  operators' best responses, i.e., the operators make offers about the amount of spectrum to contribute using a sequential updating procedure until reaching consensus. Besides spectrum allocation, we need a mode selection scheme for the multi-operator D2D users. We use a stochastic geometry framework to capture the impact of mode selection on the distribution of D2D users and assess the performance of the best response iteration algorithm. With the performance metrics considered in the paper, we show that the best response iteration has a unique Nash equilibrium that can be reached from any initial strategy. In general, asymmetric operators would contribute unequal amounts of spectrum for multi-operator D2D communication. Provided that the multi-operator D2D density is not negligible, we show that both operators may experience significant performance gains as compared to the scheme without spectrum sharing.
\end{abstract}

\section{Introduction}
\label{sec:Introduction}
Device-to-device (D2D) communication refers to the situation where two user equipments bypass the cellular \ac{BS} and core network and establish a direct local communication link. D2D communication has been introduced as one of the highlights of 3GPP Release 12~\cite{3GPP2012,3GPP2013,3GPP2014}. It offers several advantages, e.g., higher transmission rates due to proximity, reduced battery consumption at the mobile handsets, reuse and hop gain, etc.~\cite{Doppler2009,Fodor2012}. Along with benefits, D2D support brings also new challenges because the cellular network needs to cope with new interference situations, e.g., cross-tier interference between cellular and D2D users~\cite{Fodor2012}. 

In order to alleviate the interference problems introduced by D2D, a great deal of mode selection and resource allocation algorithms have been proposed, see for instance~\cite{Yu2011,Lin2013,Cho2014}. Unfortunately, almost all existing algorithms are applicable to single-operator networks. However, without multi-operator support, i.e., when the two ends in the D2D pair have subscriptions with different \acp{MNO}, the business potential of commercial D2D communication would be very limited.

The only available studies for multi-operator D2D can be found in~\cite{Phan2011,Pais2014}. Both patents designed protocols to setup a D2D communication session considering different \acp{MNO}. However, they are not seen to address how the spectrum is allocated, which spectrum is used and how the communication mode is selected. In this paper, we assume that multi-operator D2D discovery has been handled, using for instance the protocol in~\cite{Phan2011}, and we propose algorithms for multi-operator D2D spectrum allocation and mode selection. 

D2D communication can be enabled either over licensed or unlicensed spectrum. D2D communication in unlicensed bands would suffer from unpredictable interference. Also it poses a requirement for two wireless communication interfaces and efficient power management at the D2D users~\cite{Asadi2014}. Due to these reasons, at this moment, licensed spectrum seems to be the way forward to enable D2D communication, especially considering safety related scenarios such as vehicle-to-vehicle communication. In the cellular band, either dedicated spectrum can be allocated to the D2D users (a.k.a. D2D overlay) or D2D and cellular users can be allocated over the same resources (a.k.a. D2D underlay). In a multi-operator D2D underlay, cellular users may suffer from inter-operator interference, and in order to resolve it, information exchange between the \acp{MNO} might be needed. Due to the fact that \acp{MNO} may not be willing to reveal proprietary information to the competitors, we believe that, at a first stage, the overlay multi-operator D2D scheme would be easier to implement. 

In this paper, we consider a scenario with two \acp{MNO} and we identify how much spectrum each \ac{MNO} should commit for overlay multi-operator D2D communication. The \acp{MNO} need not contribute equal amount of spectrum as they may have different network utilities and loads. We formulate a game where the \acp{MNO} make offers about the amount of spectrum they want to contribute and use a sequential best response to each other. The \acp{MNO} neither need to exchange proprietary information nor to communicate with an external party. Also, they are not forced to take any action. If one party does not experience performance gain as compared to no sharing, the agreement can break and the communication among the devices is routed to the cellular infrastructure. 

Besides spectrum allocation, the \acp{MNO} also need to agree about the mode selection scheme for the multi-operator D2D users. In principle, existing schemes for single-operator networks enabling D2D communication can be used. We consider the algorithm proposed in~\cite{Cho2014} where D2D communication mode is used only if the measured interference in the part of the cellular  spectrum dedicated for D2D communication is low. We extend this algorithm to a multi-operator setting. Given the mode selection scheme, we allocate spectrum for multi-operator D2D users for maximizing the D2D user rate while also meeting \ac{MNO}-specific user rate constraints. Under these conditions, it turns out that the formulated game is concave, satisfies the dominance solvability condition, and is also sub-modular. As a result, one can prove that the best response optimization converges monotonously to the unique \ac{NE} from any initial point. 

Using the proposed scheme we illustrate that both \acp{MNO} may experience significant performance gains as compared to no sharing. To the best of our knowledge, this is the first work which uses a non-cooperative game theoretical model for spectrum allocation for multi-operator D2D communication.
\begin{figure}[t!]
\centering
\includegraphics[scale=0.19]{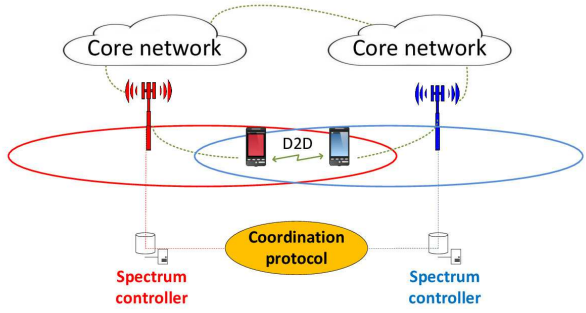}
\caption{User equipments registered to different \acp{MNO} communicating in a D2D manner.}   
\label{fig:MultiOP}
\end{figure}

\section{System model}
\label{sec:Architecture}
We consider two \acp{MNO}, $i=\{1,2\}$, enabling D2D communication. For the $i$-th operator, the \acp{BS} are distributed according to a \ac{PPP} with density $\lambda_i^b$ and, the cellular users are distributed according to a \ac{PPP} with density $\lambda_i^c$. The distributions of \acp{BS} and cellular users are independent. A cellular user is associated with the nearest home-operator \ac{BS}. An \ac{MNO} has two types of D2D users: intra-operator D2D users, i.e., when the two ends in the D2D pair have subscriptions with the same \ac{MNO}, and multi-operator D2D users. The locations of intra-operator D2D transmitters follow a \ac{PPP} with density $\lambda_i^d$ and the locations of multi-operator D2D transmitters follow a \ac{PPP} with density $\lambda$. For the multi-operator D2D pairs, we assume that the densities of transmitters from different \acp{MNO} are equal, i.e., $\lambda/2$. As a result, for the $i$-th \ac{MNO}, the ratio of multi-operator D2D to intra-operator D2D transmitters is $w_i\!=\!\lambda/(\lambda + 2\lambda_i^d)$. 

Figure~\ref{fig:MultiOP} illustrates the D2D concept towards a multi-operator scenario. Devices registered to different \acp{MNO} can communicate in a D2D manner or in infrastructure mode via the nearby serving \acp{BS}. The spectrum controller is an entity to coordinate the agreement about spectrum allocation and mode selection between the peer networks.

\begin{figure}[t!]
\centering
\includegraphics[scale=0.5]{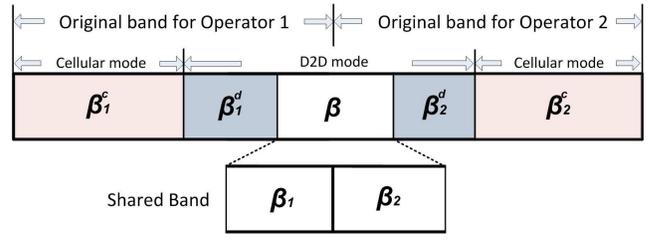}
\caption{Spectrum divisions for the \acp{MNO}.}   
\label{fig:Overlay}
\end{figure}

Figure~\ref{fig:Overlay} shows the  spectrum allocation for the \acp{MNO} in case they employ the overlay principle also for  intra-operator D2D communication. A fraction $\beta_i^c$ of the \ac{MNO}'s spectrum is dedicated for cellular communication and a fraction $\beta_i^d$ is dedicated for intra-operator D2D communication. Finally, an \ac{MNO} contributes a fraction $\beta_i$ of spectrum to the shared band, $\sum_i \beta_i$, where multi-operator D2D communication takes place.  Obviously, $\beta_i^d \!+\! \beta_i^c \!+\! \beta_i \!=\! 1, \forall i$. D2D users selecting cellular transmission mode would be allocated to the $\beta_i^c$ part of the spectrum. While in our analysis we assume FDD \acp{MNO} that contribute frequency resources for D2D communication, the same analysis is applicable to TDD \acp{MNO} that contribute time-frequency resource blocks. In the TDD case, multi-operator D2D support poses a requirement for time synchronization between the \acp{MNO} which is more challenging. 

In order to describe the quality-of-service offered to the D2D users, we assume that an \ac{MNO} maintains a network utility function that is equal to the average D2D rate 
\begin{equation}
\label{eq:Ui}
U_i = (1-w_i) \,Q_i^d + w_i \,Q_i^s, \,\, i\!=\! \{1,2\}
\end{equation} 
where $Q_i^d$ is the average normalized rate for intra-operator D2D users and $Q_i^s$ describes the same quantity for  multi-operator D2D users. Also, an \ac{MNO} must offer to its cellular users an average rate equal to a target rate, i.e., $Q_i^c \!=\! \tau_i.$

The average rate of cellular users, $Q_i^c$, can be obtained by scaling their average spectral efficiency, $R_i^c$, with the normalized bandwidth available for cellular transmissions, $\beta_i^c$. On the other hand, the D2D users may operate either in cellular or in D2D mode. Let us denote by $q_i^d$ the fraction of intra-operator D2D users selecting D2D mode and by $q$ the same quantity for multi-operator D2D users. The average rate for intra-operator D2D users, $Q_i^d$, should be computed as an average of their average spectral efficiencies in cellular mode, $R_i^c$, and D2D mode, $R_i^d$, scaled with the appropriate fractions of user density and cellular bandwidth. The average rate for multi-operator D2D users, $Q_i^s$, should be obtained in a similar manner. To sum up, 
\begin{equation}
\begin{array}{*{20}l}
Q_i^c &=& \beta_i^c  \, R_i^c \\
Q_i^d &=& \beta_i^c  \, R_i^c  (1-q_i^d) + \beta_i^d \,  R_i^d  \, q_i^d\\
Q_i^s &=& \beta_i^c  \, R_i^c  (1-q) + \beta \, R \,  q
\end{array}
\end{equation}
where $R$ is the average spectral efficiency for multi-operator D2D users selecting D2D mode. The average spectral efficiencies for cellular and D2D users can be calculated as in~\cite{Lin2013}
\begin{equation}
\begin{array}{*{20}l}
R_i^c  &=& \nu_i\int\nolimits_0^\infty{\frac{\mathcal{P}_i^c}{1+\gamma} d\gamma} \\ 
R_i^d &=& \int\nolimits_0^\infty{\frac{\mathcal{P}_i^d}{1+\gamma} d\gamma} \\ 
R &=& \int\nolimits_0^\infty{\frac{\mathcal{P}}{1+\gamma} d\gamma}
\end{array}
\end{equation}
where $\gamma$ is the SINR, $\mathcal{P}$ is the coverage probability for multi-operator D2D users in D2D mode, i.e., the probability that the SINR at a typical multi-operator D2D user is larger than the SINR $\gamma$, $\mathcal{P}_i^d$ describes the same quantity for intra-operator D2D users, $\mathcal{P}_i^c$ is the coverage probability for cellular users and $\nu_i$ is the portion of time a user in cellular mode is active. 

The coverage probability depends on the interference level at the user. The density of interferers can be computed only after the mode selection scheme is specified. There are many schemes available in the literature, e.g., based on the D2D pair distance~\cite{Lin2013} and/or the distance between the D2D transmitter and cellular \ac{BS}~\cite{Yu2011}. In these cases, D2D pairs can be arbitrarily close to each other. In order to avoid that, in~\cite{Cho2014}, it is proposed to select the mode based on the measured interference at the D2D transmitter. When the measured interference is below a threshold, there is indication there are few ongoing D2D communication pairs. Provided that the D2D pair distance $d$ is small, low interference at the D2D transmitter necessitates low interference at the D2D receiver. 

It is straightforward to extend the mode selection algorithm described in~\cite{Cho2014} in a multi-operator setting. The D2D transmitter measures the interference over the shared band, $\sum\nolimits_i{\beta_i}$, and communicates a quantized version of it to its home \ac{BS}. The \ac{BS} decides about the mode using the same threshold-based test and communicates its decision back to the transmitter. In this paper, the measurement is assumed to be done at the D2D transmitter in order to simplify the mathematical analysis of outage probability but practical implementation could be based on measurements conducted by the D2D receiver. In that case, the mode would be selected at the home operator of the receiver. As long as the D2D pair distance $d$ is short, we do not expect significant differences in the results. 

In~\cite{Cho2014}, it is shown that the locations of D2D transmitters selecting D2D communication mode follow a \ac{MPP} type II with hardcore distance $\delta$. Using the properties of MPP type II, the hardcore distance can be mapped to the mode selection threshold, $\epsilon$, using simplified methods as in~\cite{Cho2013}. Finally, the coverage probability for multi-operator D2D users in the presence of Rayleigh fading can be computed using a similar approach as in~\cite{Cho2014}  
\begin{equation}
\label{eq:cov}
\mathcal{P} \!=\! {e}^{-\!\frac{\gamma \sigma^2 \beta }{P_t l(d) } - q \lambda  \int\limits_0^{2\pi}\int\limits_{2\delta}^{\infty}
  \frac{f(r)r}{1+f(r)} {d}r {d}\phi - {c q \lambda}
  \int\limits_0^{2\pi}\int\limits_{\delta}^{2\delta}
  \frac{f(r)r}{1+f(r)} {d}r{d}\phi}
\end{equation}
where $\beta\!=\!\beta_1\!+\!\beta_2$, $P_t$ is the D2D transmit power level, $\sigma^2$ is the noise level calculated over the full cellular band, $c\!=\! 2\pi (4\pi/3 \!+\! \sqrt{3}/2)^{-1}$, $l(\cdot)$ is the distance-based pathloss and $f(r)\!=\! \gamma\, l(\sqrt{r^2\!+\!{d}^2\!-\!2rd \cos{\phi}})/l(d)$. 

The coverage probability for intra-operator D2D users can be expressed in a form similar to equation~\eqref{eq:cov} after replacing $q$ by $q_i^d$, $\lambda$ by $\lambda_i^d$ and $\beta$ by $\beta_i^d$. Finally, the coverage probability for the cellular  uplink has been derived in~\cite{Cho2014}. If a power law model for the distance-based pathloss is used, with pathloss exponent $a$, the coverage probability can be written as 
\begin{equation}
\begin{array}{*{20}l}
\mathcal{P}_i^c = \left( {1 \!+\! \alpha_i \frac{2\gamma}{a\!-\!2}\,\,  {}_2F_1\left(1,\frac{a\!-\!2}{a},2\!-\!\frac{2}{a},-\gamma\right)}\right)^{-1} 
\end{array}
\end{equation}
where $\alpha_i$ is the probability a \ac{BS} is active and ${}_2F_1$ is the Gaussian hypergeometric function. Note that the activity probability $\alpha_i$ should take into account not only the densities of cellular users but also the densities of intra-operator D2D and multi-operator D2D users selecting cellular communication mode, i.e., $(1-q_i^d)\lambda_i^d$ and $(1-q)\lambda/2$ respectively. 

\section{Multi-Operator D2D Spectrum Sharing Game}
\label{sec:Multi}
We consider a strategic non-cooperative spectrum sharing game between two \acp{MNO}, $\mathcal{G} = (\mathcal{I}, \mathcal{S}, \mathcal{U})$, where $\mathcal{I}$ is the set of \acp{MNO}, $\mathcal{S}\!=\! S_1 \!\times\! S_2$ is the set of the joint strategies  and ${\mathcal{U}} \!=\! [U_1, U_2]$ is the vector of utility functions. The strategy space for an \ac{MNO} represents the spectrum fraction contributed for the shared band, i.e., $S_i \!=\! \{\beta_i \!:\ 0 \leq \beta_i \leq u_i \},\, i\!=\!\{1,2\}$. The upper limit of the strategy space, $u_i$, depends on the \ac{MNO}-specific constraints which are presented below.  

In this paper, we assume that the mode selection threshold, $\epsilon^d$, for the spectrum band dedicated to intra-operator D2D has been decided by each \ac{MNO}, and the mode selection threshold, $\epsilon$, for the shared band dedicated to multi-operator D2D  has been agreed between the \acp{MNO} or imposed by the regulator. Note that given the decision thresholds, $\epsilon^d$ and $\epsilon$, an \ac{MNO} is able to compute the densities of intra-operator D2D and multi-operator D2D users selecting cellular communication mode.

In a non-cooperative game, each player sets its strategy profile to maximize its own utility function. For the game in question, an \ac{MNO} maximizes the average D2D user rate, see equation~\eqref{eq:Ui}, under operator-specific constraints for cellular users  and intra-operator D2D users. For instance, the average cellular user rate should be equal to a target rate, $Q_i^c \!=\!\tau_i$, and the intra-operator D2D rate should be higher than a constraint, $\beta_i^d\, R_i^d \!\geq\!\mu_i^d$. We assume that without spectrum sharing, i.e, $\beta_i\!=\!0$, these constraints are satisfied. To sum up, an \ac{MNO} will identify the amount of spectrum to contribute for multi-operator D2D communication, $\beta_i$, as the solution of the following optimization problem. 
\begin{IEEEeqnarray}{lcc} 
\label{eq:P1}
  \mathop {{\rm{Maximize:}}}\limits_{\beta_i\geq 0} &
  { U_i. } \IEEEyessubnumber\label{eq:P1:a}\\  
  {{\rm{Subject}}\,\,{\rm{to:}}} & {Q_i^c = \tau_i}\IEEEyessubnumber \label{eq:P1:b}\\
	& {\beta_i^d \, R_i^d \geq \mu_i^d.}\IEEEyessubnumber\label{eq:P1:c}
\end{IEEEeqnarray}

Since the inter-operator spectrum sharing game should be distributed and non-cooperative, one possible way to reach to a consensus is the best response iteration. According to it, given the opponent's proposal $\beta_j$, the $i$-th \ac{MNO} identifies its contribution $\beta_i$ for maximizing $U_i$. In order to identify the optimal spectrum fraction $\beta_i$ we use the following  steps: Given the thresholds $\epsilon_i^d$ and $\epsilon$, the $i$-th \ac{MNO} computes the fractions of intra-operator D2D and multi-operator D2D users in cellular communication mode, i.e.,  $(1\!-\!q_i^d)\lambda_i^d$ and $(1\!-\!q)\lambda/2$ respectively, thus able to identify the required amount of spectrum for cellular users, $\beta_i^c$, based on the equality constraint~\eqref{eq:P1:b}. Then, the spectrum fractions, $\beta_i$ and $\beta^d_i$, can be related as $\beta_i \!= \! 1 \!-\!\beta_i^c\!-\! \beta_i^d$. In the Appendix, we show that the utility and the left-hand side of the constraint in~\eqref{eq:P1:c} are both concave in $\beta_i$. If there is always a $\beta_i$ such that the constraint in~\eqref{eq:P1:c} is strictly satisfied, the first-order conditions are both necessary and sufficient. As a result, the optimal $\beta_i$ can be identified using standard convex optimization tools at a low complexity.

In a non-cooperative game, some of the most important questions are the existence and uniqueness of the \ac{NE}. In case there are multiple Nash equilibria, the selected equilibrium would depend on the initial strategy profile~\cite{Rosen, Yao}. This property might be undesirable because the performance of an \ac{MNO} would depend on the selection order between the \acp{MNO} and also on their initial proposals for spectrum contribution in the shared band. 

\begin{proposition} The formulated game has a unique \ac{NE} and the best response iteration converges to it from any initial point.
 \label{prop3}
 \begin{proof}
According to~\cite[Theorem 1]{Rosen}, a \ac{NE} exists for a concave game. In \cite[Theorem 4.1]{Moulin1980}, the dominance solvability condition under which a concave game has a unique \ac{NE} and the best response iteration converges to it from any initial point is proposed. In the Appendix, we prove that the game is concave and the utility function $U_i$ subjecting to \ac{MNO}-specific constraints satisfies the dominance solvability condition.
\end{proof}
\end{proposition}

\begin{figure}[t]
\centering
\includegraphics[scale=.56]{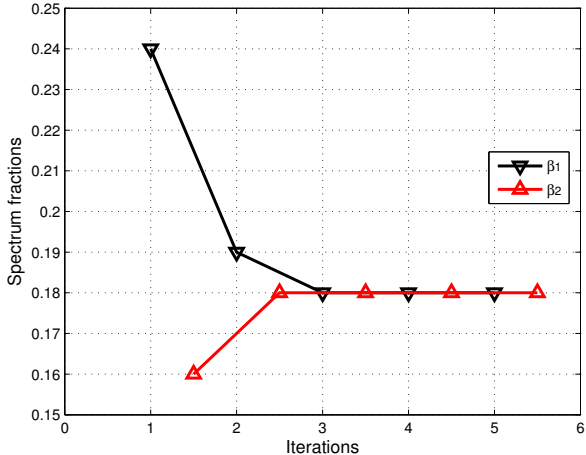}
\caption{Example convergence to the unique \ac{NE} for symmetric operators, when the mode selection threshold in the shared band is $\epsilon\!=\!-72$ dBm and in the intra-operator D2D band is $\epsilon_i^d\!=\!-75$ dBm for both \acp{MNO}.}
\label{fig:BR}
\end{figure}

\begin{proposition} Using the best response iteration, the \acp{MNO} converge monotonously to the \ac{NE}.

\label{prop1}
\begin{proof}
According to~\cite[Algorithm I]{Yao}, the best response algorithm converges monotonically to a \ac{NE}, from any initial strategy profile, if the utility $U_i$ is sub-modular function and the strategy space satisfies the descending property. The monotonicity would be in the opposite direction for the \acp{MNO}. According to~\cite{Altman}, the function $U_i$ is sub-modular in the strategy set of the two players if the first-order cross derivative is negative. Following an approach similar to the one followed in equations~\eqref{eq:prop3:1} and~\eqref{eq:prop3:2}, one can show that $\frac{\partial^2 U_i}{\partial \beta_i \partial \beta_j}\!<\!0$. Also, due to the fact that the strategy space does not depend on the opponent behavior, but only on the operator-specific constraints, the descending property is satisfied.
\end{proof}
\end{proposition}


After the best response converges, it is natural to assume that the agreement will break if the utility of an \ac{MNO} is lower than the utility corresponding to no spectrum sharing, $U_i \!<\! U_{0,i}$. 

In general, the \acp{MNO} may have different network utility functions and constraints. Because of that, the best response iteration cannot be used to infer information about the network state of the opponent \ac{MNO}.

\section{Numerical illustration}
\label{sec:Numerical}
We consider two \acp{MNO} with BS density $\lambda_i^b\!=\!1/(\pi \, 200^2),\,i\!=\!\{1,2\}$, cellular user density  $\lambda_i^c\!=\lambda_i^b$ and multi-operator D2D density, $\lambda\!=\!4\lambda_i^b$. We evaluate the performance of the spectrum allocation scheme for different intra-operator D2D densities. We fix $\lambda_1^d\!=\!\lambda_i^b$ and we vary  $\lambda_2^d$. We take 3GPP propagation environment~\cite{3GPP} into account with pathloss equation in dB: $37.6 \log_{10}(r) \!+\! 15.3$ for the cellular mode and, $40.0\log_{10}(r) \!+\! 28$ for the D2D mode, where $r$ is the distance in meters. The D2D link distance is fixed to $d\!=\!30$ m. We use fixed transmit power levels equal to $23$ dBm for the cellular mode and $20$ dBm for the D2D mode. The normalized target rates for intra-operator D2D users and cellular users are $\tau_i\!=\mu_i^d\!=\!0.3,\, i\!=\!\{1,2\}$. 

Figure~\ref{fig:BR} shows an example convergence of the best response iteration to the unique \ac{NE} for $\lambda_1^d \! = \! \lambda_2^d$. If the \acp{MNO} are symmetric, they should contribute equal amounts of spectrum in the shared band. The initial strategy profiles are randomly selected. The \ac{MNO} $1$ first solves the optimization problem~\eqref{eq:P1} assuming that the \ac{MNO} $2$ does not contribute any spectrum in the shared band. Then, the \ac{MNO} $2$ optimizes for $\beta_2$ given that the \ac{MNO} $1$ contributes spectrum fraction $\beta_1\!=\!0.24$ and so forth. According to Proposition 2, one operator converges monotonously increasing and the other monotonously decreasing to the \ac{NE}.  
\begin{figure}[t]
\centering
\includegraphics[scale=0.56]{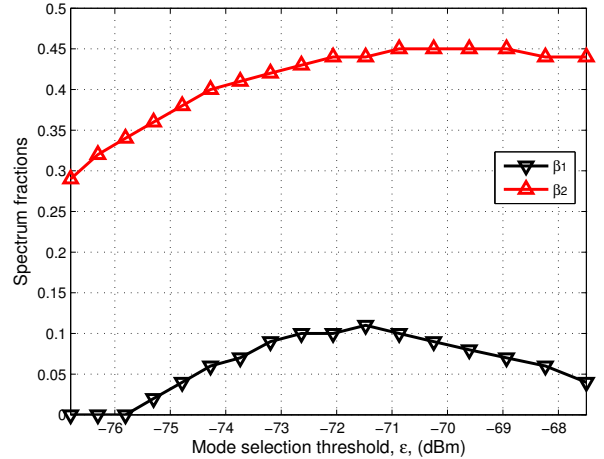}
\caption{Spectrum fractions, $\beta_i$, for multi-operator D2D w.r.t the mode selection threshold in the shared band. $\lambda_1^d\!=\!1, \lambda_2^d\!=\!0.8$. The mode selection threshold in the intra-operator D2D band is  $\epsilon_i^d\!=\!-75$ dBm for both \acp{MNO}.} 
\label{fig:beta}
\end{figure}

Next, we make the operators non-symmetric assuming that the \ac{MNO} $2$ has less intra-operator D2D users, $\lambda_2^d\!=\!0.8$. In Figure~\ref{fig:beta}, we depict the spectrum fractions contributed by the \acp{MNO} with respect to the mode selection threshold in the shared band. In general, asymmetric \acp{MNO} would contribute unequal amounts of spectrum. In our example, the \ac{MNO} $2$ has less network load than \ac{MNO} $1$ and, because of that, it has the capacity to contribute more spectrum in the shared band. Also, one can point out that a low mode selection threshold results in more users in cellular communication mode and thus, more spectrum resources should be reserved by the \acp{MNO} to meet their cellular user rate constraints. As the mode selection threshold increases, the associated bandwidth for multi-operator D2D support increases too. However, increasing the mode selection threshold beyond certain point has adverse effects, since the increased self-interference among the multi-operator D2D pairs starts reducing their rate performance. 

\begin{figure}[t]
\centering
\includegraphics[scale=0.56]{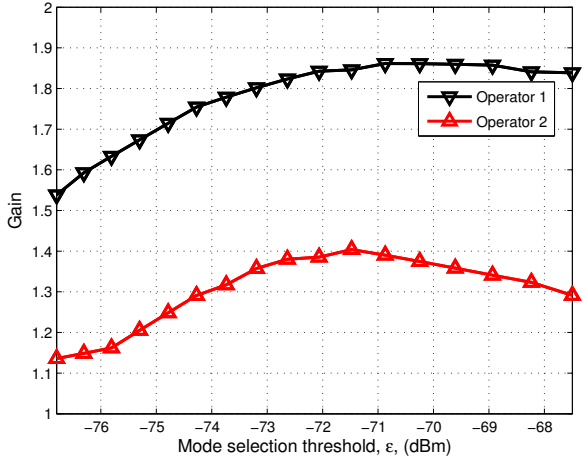}
\caption{Performance gain for the \acp{MNO} as compared to the case without spectrum sharing. $\lambda_1^d\!=\!1$,  $\lambda_2^d\!=\!0.8$, The mode selection threshold in the intra-operator D2D band is $\epsilon_i^d\!=\!-75$ dBm for both \acp{MNO}.}  
\label{fig:gain}
\end{figure}

Figure~\ref{fig:gain} shows the sum rate gain for the \acp{MNO} as compared to the case without multi-operator D2D support, i.e., all multi-operator transmissions are routed through the \acp{BS}. Both \acp{MNO} experience performance gain. The \ac{MNO} $2$ that has less network load contributes the higher fraction of spectrum in the shared band, see Figure~\ref{fig:beta}. Because of that, the \ac{MNO} $1$ enjoys more benefit from spectrum sharing than the \ac{MNO} $2$. 

\begin{figure}[t]
\centering
\includegraphics[scale=0.56]{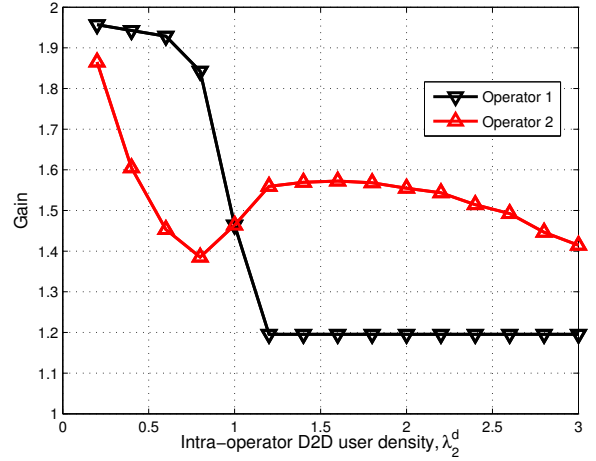}
\caption{Performance gain for the \acp{MNO} as compared to the case without spectrum sharing for varying density of D2D users for the MNO $2$. $\lambda_1^d\!=\!1$. The mode selection threshold in the shared band is $\epsilon\!=\!-72$ dBm and in the intra-operator D2D band is $\epsilon_i^d\!=\!-75$ dBm for both \acp{MNO}.}
\label{fig:gain2}
\end{figure}

Finally, we fix the mode selection threshold in the shared band, $\epsilon\!=\!-72$ dBm, and assess the performance gains for varying density of D2D users for the \ac{MNO} $2$. In Figure~\ref{fig:gain2}, one can see that symmetric \acp{MNO} achieve around $50$ \% gain. For densities $\lambda_2^d\!>\!1.2$, the \ac{MNO} $2$ does not have the capacity to contribute any spectrum, however, the \ac{MNO} $1$ still benefits by contributing a fraction $\beta_1\!=\!0.24$. In that case, the \ac{MNO} $1$ experiences $20 \%$ gain due to the proximity between multi-operator D2D users, while the gain for the \ac{MNO} $2$ is attributed to both proximity and spectrum sharing. The performance gain for both \acp{MNO} is high, i.e., close to $100 \%$, only if the network load for the \ac{MNO} $2$ becomes low. In that case, the \ac{MNO} $2$ is able to contribute a high bandwidth fraction $\beta_2 \!=\! 0.52$, while for the \ac{MNO} $1$, $\beta_1\!=\!0.08$. 

\section{Conclusions}
\label{sec:Conclusions}
In this paper, we proposed a method for spectrum allocation for D2D communication considering different mobile network operators. In a multi-operator D2D setting, the operators may not be willing to reveal proprietary information to the competitor and/or to other parties. Because of that, we modeled their interaction as a non-cooperative game. An operator makes an offer about the amount of spectrum to contribute for multi-operator D2D communication considering only its individual performance. While making the offer, it also takes into account the offer made by the opponent operator. With the performance metrics considered in the paper, we showed that the formulated game has a unique \ac{NE} and the sequence of operators' best responses converges to it from any initial point. In general, asymmetric operators contribute unequal amount of spectrum. An operator may not contribute any spectrum at all but still, the opponent may have the incentive to be cooperative due to the D2D proximity gain. Provided that the multi-operator D2D density is not negligible, we showed that both operators may experience significant performance gains. The particular gain would depend on the operator-specific network load, utility and design constraints. As potential directions for future work, one may consider spectrum sharing for more than two \acp{MNO} enabling multi-operator D2D communication. One could study whether it is beneficial to construct a common pool of spectral resources or to realize multi-operator D2D by means of bilateral agreements between operators.

\section*{appendix}
First, we show that the utility $U_i$ is concave in $\beta_i$ by computing its second derivative 
\begin{equation}
\label{eq:prop3:1}
\begin{array}{*{5}l}
\frac{\partial^2 U_i}{{\partial\beta_i}^2}&\!\!\!\!\!=(1-w_i) \frac{ \partial^2 Q_i^d}{{\partial\beta_i}^2} + w_i \frac{ \partial^2 Q_i^s}{{\partial\beta_i}^2} \\
&\!\!\!\!\!=\frac{(1-w_i) q_i^d}{\eta} \! {\displaystyle \int\limits_0^{\infty}} \frac{e^{-\frac{\gamma {\beta_i^d}}{\eta} - C(\gamma,q_i^d)}}{1+\gamma}  \left(\frac{\beta_i^d \gamma}{\eta} \!-\! 2\right)\!\! \gamma\mbox{d}\gamma \\
&+ \frac{w_i q}{\eta} \! {\displaystyle \int\limits_0^{\infty}}\frac{e^{-\!\frac{\gamma ({\beta_i\!+\!\beta_j})}{\eta} - C(\gamma,q)}}{1+\gamma}  \!\!\left(\!\frac{({\beta_i \!+\! {\beta_j}})\gamma}{\eta}\! -\! 2\!\right)\!\!\gamma \mbox{d}\gamma 
\end{array}
\end{equation}
where $\!C(\gamma,q)\!\!\!=\!\! q \lambda \!\int\limits_0^{2\pi}\! \int\limits_{2\delta}^{\infty} \!\frac{f(r)r}{1\!+\!f(r)} {d}r{d}\phi \!+\! cq \lambda\! \int\limits_0^{2\pi}\!\int\limits_{\delta}^{2\delta}\! \frac{f(r)r}{1\!+\!f(r)} {d}r{d}\phi$, $C(\gamma,q_i^d)$ can be expressed in a similar manner and $\eta \!=\! \frac{P_tl(d)}{\sigma^2}$. 

The first part of the right-hand side of the equation~\eqref{eq:prop3:1} can be treated as follow 
\begin{equation}
\label{eq:prop3:2} 
\begin{array}{*{10}l} 
{}&\!\!\!\!\!\!\!\!\!\!\!\!\!\!\!\!\!{\displaystyle \int_0^{\infty}} \frac{e^{-\frac{\gamma {\beta_i^d}}{\eta} - C(\gamma,q_i^d)}}{1+\gamma} \left(\frac{\beta_i^d \gamma}{\eta} - 2\right) \gamma \mbox{d}\gamma  \\
{}&\!\!\overset{(p1)}{\leq} {\displaystyle{\int_0^{\infty}}} \frac{e^{-\frac{\gamma {\beta_i^d}}{\eta}}}{1+\gamma} \left(\frac{\beta_i^d \gamma}{\eta} - 2\right) \gamma \mbox{d}\gamma  \\
{}&\!\!\overset{(p2)}{=}  - \frac{\rho+1}{\rho}  + (\rho+2) e^{\rho_1}{E}_1(\rho) \\
{}&\!\! \overset{(p3)}{<}  - \frac{\rho+1}{\rho}  + (\rho+2) \left( \frac{1}{\rho + ~} \frac{1}{1+~ }\frac{1}{\rho }\right) \overset{}{=} 0,
\end{array}
\end{equation}
where  $\rho \! = \!\frac{{\beta_i^d} }{\eta}$, inequality ($p1$) holds true due to $C(\gamma,q_i^d)\!\!>\!\!0$ and equality ($p2$) uses that $\int_0^{\infty}  \frac{e^{- \rho x }}{1+x} \left(\rho x- 2 \right) x~dx \!=\! - \frac{\rho+1}{\rho} \! +\! (\rho+2) e^{\rho}{E}_1(\rho)$ where $E_1(\rho) \!=\! \int_{\rho}^{\infty} \frac{e^ {-\gamma}}{\gamma} \mbox{d}\gamma$ is the exponential integral. For $\rho \!>\! 0$, there is a continued fraction form expressed as $E_1(\rho) \!\!=\!\! e^{-\rho} \left( \frac{1}{\rho + ~ } \frac{1}{1+~ } \frac{1}{\rho + ~ } \frac{2}{1+~ } \cdots \right)$ from \cite[5.1.22]{Abramowitz}. This continued fraction form is less than $e^{-\rho} \left( \frac{1}{\rho + ~ } \frac{1}{1+~} \frac{1}{\rho}\right)$. From this relation, inequality ($p3$)  holds true.

In a similar manner one can show that the second term of the right-hand side of equation~\eqref{eq:prop3:1} is also negative.  That completes the proof that the utility $U_i$ is concave. Also, the utility $U_i$ is continuous in $\beta_j$. The constraint~\eqref{eq:P1:b} just affects the upper limit of strategy space while the left-hand side of the constraint in~\eqref{eq:P1:c} is concave in $\beta_i$. That completes the proof that the game in question is concave. 

The Lagrangian function of the optimization problem~\eqref{eq:P1} is  $L_i = U_i + \nu_i(\beta_i^c R_i^c  - \tau_i^c) +  \xi_i(\beta_i^d R_i^d - \mu_i^d)$, where $\nu_i\geq 0$ and $\xi_i\geq 0$ are the Lagrange multipliers. Then, the convergence of the best response to the unique NE can be proved by the diagonal dominance solvability condition of the Lagrangian.
\begin{equation} \label{eq:prop3:3} 
\begin{array}{*{10}l} 
 \left|\frac{\partial^2 L_i}{\partial\beta_i^2}\right|> \left|\frac{\partial^2 L_i}{\partial\beta_i \partial\beta_j}\right|, \forall i
\end{array}
\end{equation}
where
\begin{equation}
\label{eq:prop3:4} 
\begin{array}{*{10}l} 
\frac{ \partial^2 L_i}{\partial\beta_i^2} \!= \!
(1\!-\!w_i) \frac{\partial^2 Q_i^d}{{\partial\beta_i}^2} \!+\! w_i \frac{ \partial^2 Q_i^s}{{\partial\beta_i}^2} \!+\! \frac{ \partial^2 \xi_i (\beta_i^d R_i^d \!-\! \mu_i^d)}{\partial\beta_i^2}
\end{array}
\end{equation}
and 
\begin{equation}
\label{eq:prop3:5} 
\begin{array}{*{10}l} 
\frac{\partial^2\! L_i}{{\partial\beta_i}{\partial\beta_j}} \!=\! \frac{w_iq}{\eta} \!{\displaystyle\int\limits_0^{\infty}} \frac{e^{-\frac{\gamma ({\beta_i\!+\!\beta_j})}{\eta} - C(\gamma,q)}}{1+\gamma}\!\! \left(\!\frac{({\beta_i\!+\!{\beta_j}})\gamma}{\eta}\!\! -\!\! 2\!\right)\!\! \gamma\mbox{d}\gamma.
\end{array}
\end{equation}

%

One can see that the second term in the right-hand side of equation~\eqref{eq:prop3:4} is negative and equal to right-hand side of equation~\eqref{eq:prop3:5}. The first and third terms in the right-hand side of equation~\eqref{eq:prop3:4} are negative too thus, inequality~\eqref{eq:prop3:3} holds. 

\section*{Acknowledgment}
This work has been performed in the framework of the FP7 project ICT
317669 METIS, which is partly funded by the European Union. Also, 
this work was supported by the Academy of Finland funded
project SMACIW under Grant no. 265040.

\end{document}